\documentstyle[epsfig, twocolumn]{esapub}

\begin{document}

\setlength{\parindent}{0pt}
\setlength{\parskip}{ 10pt plus 1pt minus 1pt}
\setlength{\hoffset}{-1.5truecm}
\setlength{\textwidth}{ 17.1truecm }
\setlength{\columnsep}{1truecm }
\setlength{\columnseprule}{0pt}
\setlength{\headheight}{12pt}
\setlength{\headsep}{20pt}
\pagestyle{esapubheadings}

\title{ON THE USE OF NONLINEAR REGULARIZATION IN INVERSE
METHODS FOR THE SOLAR TACHOCLINE PROFILE DETERMINATION}

\author{{\bf T. Corbard$^{1}$, G. Berthomieu$^{1}$,  J. Provost$^{1}$ \& 
 L. Blanc-F\'eraud$^{2}$} \vspace{2mm} \\
$^{1}$Laboratoire  Cassini, CNRS UMR 6529, OCA , BP 4229, 06304  Nice Cedex 4, France\\
$^{2}$Projet Ariana, CNRS/INRIA/UNSA, 2004 route des Lucioles, BP 93, 
     06902 Sophia Antipolis Cedex, France 
}

\maketitle

\begin{abstract}

Inversions of rotational splittings 
have shown that the surface layers and the so-called solar tachocline 
at the base of the convection zone are regions in which high radial 
gradients of the rotation rate occur.
The usual regularization methods 
 tend to smooth out every high gradients in the
solution and may not be appropriate for the study of a zone like the 
tachocline.
In this paper we use
 nonlinear 
regularization methods that are developed for
edge-preserving regularization in computed imaging (e.g. Blanc-F\'eraud et al. 1995) and 
we apply them in the helioseismic context of rotational inversions.
\end{abstract}
\section{INTRODUCTION}
%

The existence of high gradients in the  solar rotation profile near
the surface  and at the base of the convection zone 
in the so-called \emph{solar tachocline} (\cite{spiegel92}) has been revealed
 by the inversions of the rotational splittings 
(see e.g. \cite*{thompson:gong96} and 
\cite*{schou:mdi98}  for the last results).
The tachocline
represents a thin zone where the differential rotation
of the convection zone becomes rigid in the radiative
interior. It is thought to be the place from where the
solar dynamo originates and its precise structure is
an important test for angular momentum transport theories.
 More precisely, the thickness 
of the tachocline can be related to the horizontal component
of the turbulent viscosity and may be used  as an important 
observational constraint on the properties of 
the turbulence (\cite{spiegel92,gough_sekii,elliot}).  

Several works have already 
been performed to infer the fine structure of the tachocline
(\cite{charbonneau98}; \cite{kosovichev96}, 1998; 
\cite{basu97}; \cite{corbard98a};
\cite{antia98}) using both forward analysis and inverse 
techniques.
For the inverse approach,
 it may be interesting to change the  global  constraint
which tends to smooth out every high gradients in the 
solution and to find a way to preserve such zones
 in the inversion process. 
A first attempt in this direction has been carried out  (\cite{corbard98a}) by
using a nonlinear regularization term through the PP-TSVD
method (\cite{PPTSVD}).
An  investigation of the possibility to use the
  elaborate nonlinear techniques, 
 developed for edge-preserving regularization in computed imaging,
in helioseismic context is being developed by \cite*{corbard98b}.
 Here we present preliminary results  obtained by this method for
 the tachocline. 

 Section~\ref{sec:pb} briefly recalls  the relation between the
solar internal rotation and  the  helioseismic measured frequency splittings
 and presents 
the corresponding discretized inverse problem. We discuss  
in Section~\ref{sec:Reg_non_lin}  the non linear approach
of regularization in inverse techniques  
 and the computational aspects.   
In  
Section~\ref{sec:choix_reg_param}, the choice of the regularizing parameters 
for the particular case of the solar
rotation inversion and 
 the preliminary results obtained with LOWL (Tomczyck et al. 1995, 2 years of data) and  MDI (144 days, Schou et al. 1998)  data are presented. 

\section{FORMULATION OF THE PROBLEM }\label{sec:pb}


In this paper we  consider the 1D problem of inferring
the solar equatorial rotation profile $\Omega_{eq}=\Omega(r,90^\circ)$
from sectoral splittings (i.e.  $m=l$)
(e.g.  \cite{duvall84,antia96})

\begin{equation}\label{eq:int}
{\nu_{nll}-\nu_{nl0}\over l}\simeq \int_0^{R_\odot} K_{nl}(r)\
 \Omega_{eq}(r)\ dr,
\end{equation} 
where $l$, $n$, $m$ are respectively  the degree, the radial order and the azimuthal order, $r$ is the solar radius 
 and $K_{nl}(r)$ are the so-called rotational kernels that have been
calculated for each mode from a solar model taken from 
\cite*{morel97}. 

This approximation of the 2D integral equation which relates the internal rotation to the splittings, is valid only for 
high degree modes (e.g. \cite{corbard97:capo}) but the influence
of  low degree modes on the determination of  the tachocline 
and upper layers is thought
to be small. Moreover the rotation is 
known to be rigid or quasi rigid in the radiative interior.

We search a solution $\bar\Omega(r)$ as a piecewise linear function
of the radius by setting:
\begin{equation}\label{eq:expansion}
\bar\Omega(r)=\sum_{p=1}^{N_p} \omega_p\psi_p(r) \ \ \ 
  \Omega\equiv (\omega_p)_{p=1,N_p}
\end{equation}
where $\psi_p(r)$, $p=1,N_p$ are piecewise straight lines ($N_p=100$ in this
work) between fixed break points distributed according to the
density of turning points of modes (cf. \cite{Corbard97}).
The discretization of Equation~\ref{eq:int} leads to the matrix equation: 

\begin{equation}
W=R \Omega,
\end{equation}
where 
$W\equiv({W_i/\sigma_i})_{i=1,N}$ is the vector 
of the $N$ observed frequency splittings $W_i$ weighted by the standard 
deviation $\sigma_i$ given
by observers for each mode $i\equiv(n,l)$. 
No correlation between the different 
modes is assumed. 
The  matrix $R$ is  defined by: 
\begin{equation}\label{eq:discret}
R\equiv (R_{ip})_{{i=1,N\atop p=1,N_p}} \ \
R_{ip}={1\over \sigma_i}\int_0^{R_\odot} \!\!\!\!K_i(r)\psi_p(r) dr
\end{equation}

An  inverse method should lead to  a solution that 
is able to produce a good
fit of the data. We define the goodness of the fit in chi-square sense
by  the $\chi^2$ value obtained for  any solution $\bar\Omega(r)$:

\begin{equation}
\chi^2(\bar\Omega(r))=
\sum_{nl}\left[{W_i- \int_0^{R_\odot}K_i(r)
\bar\Omega(r)dr\over\sigma_i}\right]^2,
\end{equation}
which can be written in the discretized form:

\begin{equation}
\chi^2(\Omega)=\|  R \Omega-W\|_2^2.
\end{equation}

\section{ NON LINEAR REGULARIZATION}\label{sec:Reg_non_lin}
%
\subsection{ Euler equations}

The  inverse integral problem is an ill-posed problem
and the minimization of only the $\chi^2$ value 
generally leads to
oscillatory solutions that are not `physically acceptable'.
 So, a regularization technique must be used 
in the  minimization process. 
A large class of these techniques can be expressed in the general form of
the  minimization of a criterion $J$ over the unknown solution
$\bar\Omega(r)$:

\begin{equation}\label{eq:crit}
J(\bar\Omega(r))=\chi^2(\bar\Omega(r))+\lambda^2\int_0^{R_\odot} 
\varphi\left({\left| d^q\bar\Omega(r)\over dr^q\right|}\right)dr,
\end{equation}
The so-called trade-off parameter $\lambda$  is
chosen so that it  establishes  a balance between the goodness
of the fit of the data and the constraint introduced on the
solution (cf. Section~\ref{sec:choix_reg_param}).
The order $q$ of the derivative is usually taken equal to one or
two. The two choices can lead to similar results
with the appropriate choice 
$\lambda$ in the domains where the solution is well constrained
 by the data. 
In this work   we consider only smoothing term with  first derivative.

For a general $\varphi$-function, one can write  the
 criterion  Equation~\ref{eq:crit} in a discretized form:

\begin{equation}
J(\Omega)=\chi^2(\Omega)+\lambda^2 J_2(\Omega),
\end{equation}
where $J_2(\Omega)$ represents the discretized
regularization term defined by:

\begin{equation}\label{eq:J2}
 J_2(\Omega)=\int_0^{R_\odot}\!\!\!\varphi\left({\left| d\bar\Omega(r)\over dr\right|}\right)dr
=\sum_{p=1}^{N_p-1} c_p 
\varphi\left(\left|  {L}  \Omega\right|_p\right).
\end{equation}
In this equation $(c_p)_{p=1,N_p-1}$ represent the weights used for the
integration rule, $L$ is a discrete approximation of the 
first derivative operator, and $\left| L\Omega\right|_p$ is the absolute value
 of the $p$-component of the vector $L\Omega$. 
The expression of $c_p$ and $   L$
are given in  \cite*{corbard98b} for the simple case of the polynomial
expansion Equation~\ref{eq:expansion} 

The minimization of the criterion $J(  \Omega)$ leads to the following Euler equations 
(discretized form):

\begin{equation}\label{eq:euler1}
 \nabla J(\Omega)=0 \Longleftrightarrow
(R^\top R  +\lambda^2    L^\top    B(\Omega) 
 L) \Omega = R^\top   W 
\end{equation}
$B $ is a diagonal matrix. Its  elements  depend on the
gradient of the solution at each grid point: 

\begin{equation}\label{eq:euler2}
 B=diag(b_p)\ \ \ \mbox{with}\ \ \ 
b_p=c_p\times{\varphi^{'}\left(\left| {L} \Omega\right|_p\right)\over 2 
\left| {L}  \Omega\right|_p}
\end{equation}

 Two choices for the
$\varphi$-function lead to well known regularization strategies: 
\begin{itemize}

\item
$\varphi(t)=t^2$ corresponds to the traditional Tikhonov approach with first 
derivative  whereas 
\item
$\varphi(t)=t$ is known as the \emph{Total Variation} (TV) regularization
 method
(e.g. \cite{acar94}). 
It has been shown 
that this regularization method is  
able to recover piecewise smooth solutions with steep gradients 
(\cite{vogel96}).
\end{itemize}
The use of a general nonquadratic $\varphi$-function  will lead to a nonlinear 
problem which requires an appropriate iterative method to be solved. 



\subsection{Properties of the weight function ${\varphi^{'}(t)\over 2t}$}

From the Euler equation (Equations~\ref{eq:euler1} and \ref{eq:euler2})
we can see that  the function ${\varphi^{'}(t)/2t}$ acts 
as a weight function in the smoothing process: at each grid point
the gradient of the solution is used as an argument of this function
in order to set locally more or less regularization. This suggests 
an iterative process where the gradient of the solution at a given 
step is used for the computation of the regularization term at the next step.
 Three properties of the weighting function
 ${\varphi^{'}(t)/2t}$ are required to obtain a satisfactory solution and 
 to preserve high gradients 
(\cite{charbonnier97}):
\begin{enumerate}
\item 
 no smoothing for high gradients:
\begin{equation}
\lim_{t\rightarrow\infty}{\varphi^{'}(t)\over 2t}=0
\end{equation}
\item Tikhonov smoothing for low gradients:
\begin{equation}\label{eq:ppte2}
0<\lim_{t\rightarrow0}{\varphi^{'}(t)\over 2t}=M<\infty 
\end{equation}
\item 
\begin{equation}  
{\varphi^{'}(t)\over 2t} \ \ \mbox{strictly 
decreasing to avoid instabilities}. 
\end{equation}
\end{enumerate}

Either
convex or non-convex  $\varphi$-functions may be chosen 
(see  \cite*{charbonnier97} and  \cite*{teboul98} for examples in both cases). A 
non-convex function may be more suited for the search of high gradients.
But this choice leads to some numerical difficulties and instabilities
related to the existence of local minima and  may induce a high 
sensibility of the inverse process to  the regularization
parameters. 
At the opposite the choice of a convex
function  avoids these numerical problems and is more suitable for relatively
smooth transition (\cite{blanc-feraud95}).


\subsection{The iterative algorithm: ARTUR}

Following  \cite*{charbonnier97} the inversion using  
  non linear regularizing  criterion can be solved by  an iterative 
scheme named ARTUR (Algebraic Reconstruction Technic Using Regularization)
that is easy to implement:
at each step $k$ we calculate
the regularization term using the derivative of the previous 
estimate $   \Omega^{k-1}$ and  we simply
compute the new estimate $\Omega^k$  by solving the linear system:
\begin{equation}\label{eq:lin_syst}
 \Bigl(   R^\top   R  +\lambda^2    L^\top 
  {B}(   \Omega^{k-1})    L\Bigr)  \Omega^k =
   R^\top    W. 
\end{equation}
For  a convex $\varphi$-function, the convergence of this so-called 
 algorithm has been established (\cite{charbonnier97}).
This is therefore
 an adaptative regularization method which uses the information
on the derivative of the solution obtained at each step in order
to improve the regularization at the next step. This requires an
initial guess $\Omega^{0}$ for the solution but we will show in the
 next section 
that a constant solution can ever be used as the starting point.

An example of artificial discontinuous rotation has been used to test
 the algorithm. The corresponding rotational splittings have been
 computed according to Equation~\ref{eq:int}
 with addition of Gaussian noise with a standard deviation taken for each 
mode from the formal error given in observational
data (cf. \cite{corbard98a}). A second set of artificial data with the same rotational law and standard deviations divided by $\sqrt{10}$ has also been used.

 At each step of ARTUR algorithm the linear system
(Equation~\ref{eq:lin_syst}) has been solved using an iterative conjugate gradient
method 
using $\Omega^{k-1}$ as starting point. This leads to a very fast algorithm
where the number of conjugate gradient iterations needed to solve the 
linear system decreases at each ARTUR step (i.e. as $k$ increases).
The algorithm is stopped when the norm of the relative
difference between two solutions at two successive steps is  below $10^{-6}$
i.e.:
\begin{equation}
{\|\Omega^k-\Omega^{k-1}\|_2\over\|\Omega^k\|_2}\le  10^{-6}
\end{equation}

\begin{figure}[!h]
   \begin{center}
   \leavevmode
   \centerline{\epsfig{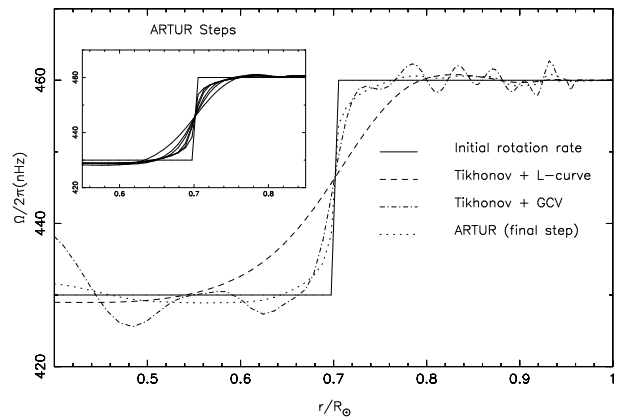}}
   \end{center}
 \caption{\em   Solutions obtained by inverting  splittings computed
from a discontinuous one dimension 
rotation profile (full line) for the same  mode set
as in LOWL data and by adding some 'realistic' noise (see text).
The standard Tikhonov
solution is given for two different automatic choices of the regularizing 
parameter. 
The successive steps of ARTUR algorithm are given in the upper left window
 whereas the final step is shown  on the main plot. 
The choice of the regularizing function and 
parameters for ARTUR  algorithm are those 
discussed  in Section ~\ref{sec:choix_reg_param} The solutions are plotted 
without  error bars for clarity. }
   \label{fig:comp.ks1}
\end{figure}

\begin{figure}[!h]
   \begin{center}
   \leavevmode
   \centerline{\epsfig{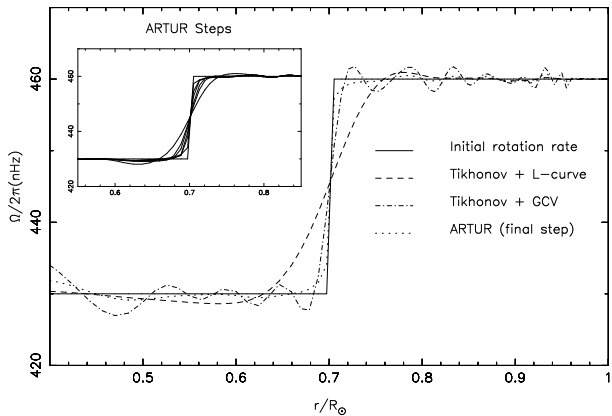}}
   \end{center}
   \caption{\em The same as Figure~\ref{fig:comp.ks1} but computed for a lower 
level of noise (standard deviations divided by $\sqrt{10}$). Comparison between Figures~\ref{fig:comp.ks1} and \ref{fig:comp.ks10} shows the smoothing effect
of the data noise level for the three methods.}
   \label{fig:comp.ks10}
\end{figure}

 The results are given in Figures~\ref{fig:comp.ks1} and \ref{fig:comp.ks10}
 which  show examples of ARTUR
steps (upper window). It is seen how the ARTUR solution, starting from a smoothed Tikhonov
 solution,  becomes steeper at each step.
Comparison between  Figures~\ref{fig:comp.ks1} and \ref{fig:comp.ks10}
 shows the effect of the errors in the data which leads to a smoothing
 of the edges of the discontinuity.

\section{APPLICATION TO THE SOLAR ROTATION 
INVERSION}\label{sec:choix_reg_param}

In the particular case  of the solar 
tachocline,
the uncertainty on the width of the transition zone is still large
(see Table 2 of \cite*{corbard98a} for a summary of some
previous works). Here we apply the non linear regularizing method to the 
LOWL and MDI sectoral splittings.

\subsection{Choice of the $\varphi$-function}

According to the previous discussion,
we have chosen to consider a  convex regularizing 
$\varphi$-function as
  \cite*{charbonnier97}: 
\begin{equation}
\varphi(t)=2\sqrt{t^2+1}-2
\end{equation}
This function is close to the absolute value function used in TV
 regularization but   has a quadratic behavior near $0$ in order to smooth 
zones with small
gradient moduli in addition to the 
 linear behavior near infinity that preserves
high gradient zones.


In addition to the usual regularizing parameter $\lambda$ 
already introduced, we have included in the $\varphi$-function
a parameter which scales the derivative of the solution. It 
 allows to adapt the domain of gradient modulus where the solution
 is very weakly smoothed to the specific problem we consider.


\begin{equation}\label{eq:critdelta}
J(\bar\Omega(r))=\chi^2(\bar\Omega(r))+\lambda^2\int_0^{R_\odot} 
\varphi\left({1\over\delta}{\left| d\bar\Omega(r)\over dr\right|}\right)dr
\end{equation}
This simply 
leads to replace $\lambda$ by  $\overline {\lambda} =\lambda/\delta$ in
 Euler Equation~\ref{eq:euler1}
and to use 
\begin{equation}\label{eq:wf}
{\varphi^{'}(t)/ 2t}={1/ \sqrt{1+\left({t/\delta}\right)^2}}
\end{equation}
where $t=\left| L  \Omega\right|_p$ 
as weighting function in Equation~\ref{eq:euler2}.

Therefore we have to define a strategy to choose the two parameters
$\overline{\lambda}$ and $\delta$.

\subsection{The choice of $\overline{\lambda}$
}
If the initial guess  $\Omega^{0}$ is  a constant function, then, according
to Equations~\ref{eq:ppte2} and \ref{eq:wf}, $M=1$ and  the solution at
the first ARTUR step corresponds to a Tikhonov solution with 
$\overline{\lambda}$  as regularizing parameter.
It has been shown in \cite*{corbard98a} that
the Generalized Cross Validation (GCV) strategy 
 leads systematically to a less
smoothed solution than the L-curve 
one (\cite{L-curve}) ($\lambda_{Lcurve}\simeq 100*\lambda_{GCV}$
in that work) 
and therefore is more suited to the study of the tachocline.
Nevertheless, this choice leads to 
spurious oscillations below and above the tachocline 
(see Figures~\ref{fig:comp.ks1}, \ref{fig:comp.ks10}).
As ARTUR algorithm will tend to enhance the high gradients found
at the first step it is important to start with a 
 solution smooth enough to  avoid spurious oscillations with high gradients.
At the opposite,
the L-curve choice leads often to a solution which is too smooth
and does not allow to exhibit the expected high gradients during the
iterations. As the optimal choice of this parameter strongly
depends on the level of noise included in the data (cf. Figures~1 and 2),
 it is important to define an automatic  choice of this parameter 
so that we use the same strategy for different datasets or for different
realizations of the noise. It consists in taking an  intermediate value
 between  the values of  $\lambda_{Lcurve}$ and
$\lambda_{GCV}$ for Tikhonov inversion. This has been used here  to fix 
 $\overline{\lambda}$ in ARTUR algorithm.

\subsection{The choice of $\delta$
}

 \begin{figure}[!h]
   \begin{center}
   \leavevmode
   \centerline{\epsfig{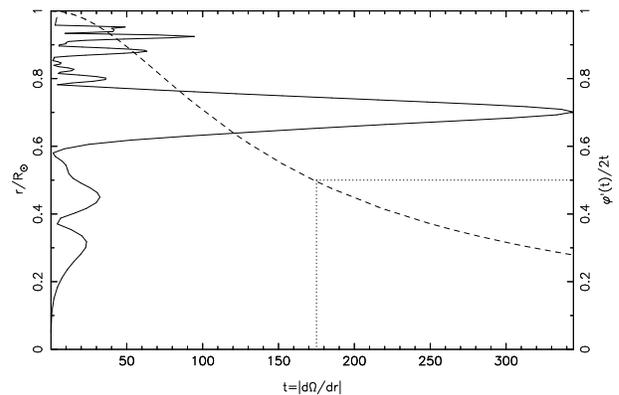}}
   \end{center}
   \caption{\em The full line shows the first derivative of a first step 
solution 
(cf. Figure~\ref{fig:comp.ks1}) in ARTUR algorithm 
as a function of the fractional solar radius. For each value 
of the gradient, the dashed line gives the weight that will be given locally 
to the regularizing term at the second step of ARTUR algorithm. The dotted 
line indicates the gradient upon which the local regularization will be 
more than 50$\%$ smaller at the second step compared to the first step.}
   \label{fig:delta}
\end{figure}				

The parameter $\delta$ is introduced to adapt the shape
of the weighting function to the gradient that we search to detect .
 Its value is chosen by looking at the derivative of the solution at the first iteration step. We have chosen for simplicity to keep
this parameter constant during the iterations.  Figure \ref{fig:delta} shows 
as an example (full line)
the first derivative of solution obtained  at the first step
by inverting artificial splittings which have been computed for the 
discontinuous 
rotation law presented in the previous section (cf. Figure~\ref{fig:comp.ks1}).
The largest peak corresponds to the rapid variation in the tachocline, the smaller ones to  spurious oscillations in the solution.
The weighting function (Equation~\ref{eq:wf})
is shown in dashed line for $\delta=100$.   According to Figure~\ref{fig:delta}, 
the choice of $\delta=100$  leads
to regularize $50\%$ less
at the second step in that zones where the gradient of the first
step solution is above $\sim 175$ nHz/$R_\odot$.

According to the previous results on 
solar rotation inversions, the width of the
tachocline does not exceed $0.1$ solar radius. This represents also
the resolution reached  at the tachocline localization 
 ($\simeq 0.69R_\odot$) with a Tikhonov method using a regularizing
parameter chosen near the corner of the L-curve. Furthermore we
have a good estimate  of the difference between the rotation rate
above and below the transition 
($\simeq 30$nHz in \cite*{corbard98a}). Therefore
we can estimate an order of magnitude  of  $300$nHz/$R_\odot$ 
for the maximum gradient obtained at the first iteration
of ARTUR process. This corresponds to the value in  Figure \ref{fig:delta}.
At the second step we want to preserve
only high gradients i.e.  to regularize less in that zones
where high gradients have already been found at the first
step.
With our a priori knowledge of the maximum gradient at the first step
($\sim 300$nHz, see also Figure~\ref{fig:delta}), the choice of $\delta = 100$ 
sounds reasonable in the sense that it will tend to decrease 
the regularization 
especially in the tachocline.
A smaller value would enhance the secondary peaks that may be induced 
by the data noise.

\subsubsection{Results for the tachocline}
 \begin{figure}[!t]
   \begin{center}
   \leavevmode
   \centerline{\epsfig{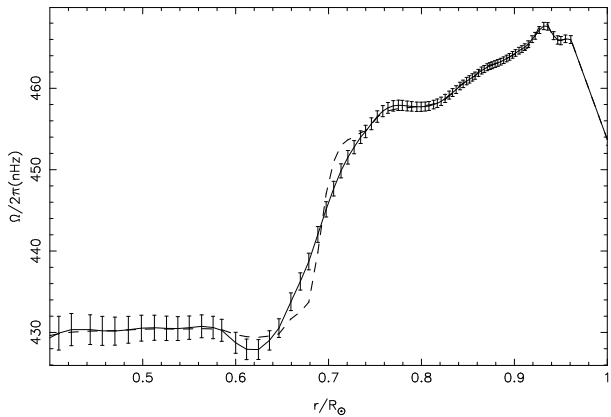}}
   \end{center}
   \caption{\em Solar equatorial rotation obtained by inverting LOWL data.
Tikhonov solution is shown by the full line with error bars. The dashed line represents the final ARTUR step.
}
   \label{fig:lowl}
\end{figure}
 \begin{figure}[!t]
   \begin{center}
   \leavevmode
   \centerline{\epsfig{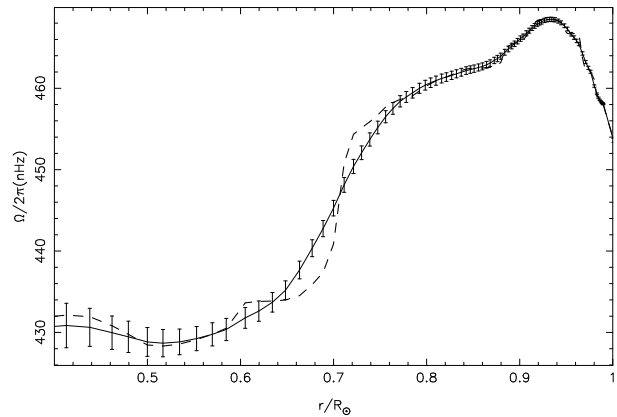}}
   \end{center}
   \caption{\em Same as Figure~5 for MDI data 
}
   \label{fig:mdi}        
\end{figure}

Figures~\ref{fig:comp.ks1} and \ref{fig:comp.ks10} show that ARTUR algorithm leads always to
 better results than Tikhonov inversion without `local deconvolution' in the 
case of a discontinuous rotation profile. Furthermore, it is shown in 
\cite*{corbard98b} that it can also lead to good results for simulated 
tachocline widths
between $0.02$ and $0.08R_\odot$.

Preliminary results have been  obtained
by inverting the sectoral splittings given by the 2 years LOWL data and the 
144 days  MDI data. The results of Tikhonov
inversion  and the last step
of ARTUR algorithm are plotted in Figures \ref{fig:lowl} and \ref{fig:mdi}. 
Tikhonov solutions are shown with error bars at each 
grid points which are deduced from the propagation of data noise 
through this linear process. ARTUR algorithm is nonlinear and therefore we 
can not compute formal errors in the same way. 

In order to have an estimate
of the uncertainties on the tachocline parameters derived by  ARTUR algorithm, 
Monte-Carlo simulations with artificial splittings have been performed by  Corbard et al. (1998b). The rotation profiles
were taken as  $erf$ functions with different 'initial widths'. Each 
'inferred width' is  the mean value  
of the results obtained by fitting  directly the solutions
by an $erf$ function, for 500 realizations of input errors.
 Error bars are estimated from  a $68.3\%$ confidence interval.
 From this study it seems that
an uncertainty of $\pm0.02R_\odot$ on the tachocline width can be reached
with both ARTUR algorithm and Tikhonov inversion with a 'local deconvolution'
using  the averaging kernels computed at the center of the transition 
(see Corbard et al. 1998b). 

The width of the tachocline estimated from LOWL data
and by using ARTUR algorithm  is $0.05\pm0.02R_\odot$ in good agreement 
with the value of $0.05\pm0.03R_\odot$ found for the same data in
\cite*{corbard98a} by studying systematically the effect of regularization
on the determination of tachocline parameters for three inverse methods.
A preliminary study of MDI data leads to a 
slightly larger
width of $0.08R_\odot$ but Monte-Carlo simulation have not yet been performed
that allows to give an estimate of the uncertainty on this value.
For both dataset the widths obtained from Tikhonov inversions after
 `local deconvolution' using 
averaging kernels are always larger ($\sim 0.1R_\odot$) than the 
estimates obtained with ARTUR algorithm.

We note that  these  estimates of the tachocline width 
 have been obtained by fitting the solutions by an $erf$ function
between $0.4$ and $0.8R_\odot$ and this may be
 better adapted to the shape of LOWL solution (Figure~\ref{fig:lowl})
 which presents a step
between $0.75$ and $0.8R_\odot$. This step is not found with 
MDI data and this may explain the larger width
found by our fit with these data. One possibility for future work is to change
our fitting function i.e. to change our definition of the tachocline
width. However we have also to explain  the
different behavior of  the two solutions near $0.8R_\odot$ and to look if this 
remains with longer time series of MDI experiment.

\section*{ACKNOWLEDGMENTS}
We acknowledge SOI team and S. Tomczyk for allowing the use of MDI and LOWL data. SoHO is a project of international cooperation between ESA and NASA.
This work has been performed using the computing facilities provided 
by the program
``Simulations Interactives et Visualisation en Astronomie et M\'ecanique''
(SIVAM, OCA, Nice) and by the ``Institut du D\'eveloppement 
et des Ressources en Informatique Scientifique'' (IDRIS, Orsay).
T. Corbard thanks the conference organizers for financial support.


\begin{thebibliography}{}

\bibitem[\protect\astroncite{Acar \& Vogel}{1994}]{acar94}
Acar, R., Vogel, C.R. 1994,
 Inverse Problems 10(6), 1217

\bibitem[\protect\astroncite{Antia et~al.}{1998}]{antia98}
Antia, H.M., Basu, S., Chitre, S.M. 1998,
 MNRAS,
 submitted

\bibitem[\protect\astroncite{Antia et~al.}{1996}]{antia96}
Antia, H.M., Chitre, S.M., Thompson, M.J. 1996,
 A\&A 308, 656


\bibitem[\protect\astroncite{Basu}{1997}]{basu97}
Basu, S. 1997,
 MNRAS 288, 572


\bibitem[\protect\astroncite{Blanc-F\'eraud et~al.}{1995}]{blanc-feraud95}
Blanc-F\'eraud, L., Charbonnier, P., Aubert, G., Barlaud, M. 1995,
 In: IEEE Proceedings of the 2nd International Conference of Image Processing,
 Washington DC, USA, p.~474

\bibitem[\protect\astroncite{Charbonneau et~al.}{1998}]{charbonneau98}
Charbonneau, P., Christensen-Dalsgaard, J., Henning, R., et~al. 1998,
 In: Provost J., Schmider F.X. (eds.) IAU Symp. 181: Sounding Solar and Stellar
  Interior (poster volume).
 OCA \& UNSA, Nice, p.~161


\bibitem[\protect\astroncite{Charbonnier et~al.}{1997}]{charbonnier97}
Charbonnier, P., Blanc-F\'eraud, L., Aubert, G., Barlaud, M. 1997,
 IEEE Trans. on Image Processing 6(2), 298

\bibitem[\protect\astroncite{Corbard}{1997}]{corbard97:capo}
Corbard, T. 1997,
 In: Proceedings of IX IRIS meeting,
 in press

\bibitem[\protect\astroncite{Corbard et~al.}{1997}]{Corbard97}
Corbard, T., Berthomieu, G., Morel, P., et~al. 1997,
 A\&A 324, 298


\bibitem[\protect\astroncite{Corbard et~al.}{1998a}]{corbard98a}
Corbard, T., Berthomieu, G., Provost, J., Morel, P. 1998a,
 A\&A 330, 1149

\bibitem[\protect\astroncite{Corbard et~al.}{1998b}]{corbard98b}
Corbard, T., Blanc-F\'eraud, L., Berthomieu, G., Provost, J. 1998b,
 A\&A, to be submitted


\bibitem[\protect\astroncite{Duvall et~al.}{1984}]{duvall84}
Duvall Jr, T.L., Dziembowski, W.A., Goode, P.R., et~al. 1984,
 Nature 310, 22

\bibitem[\protect\astroncite{Elliot}{1997}]{elliot}
Elliot, J.R. 1997,
 A\&A 327, 1222




\bibitem[\protect\astroncite{Gough \& Sekii}{1998}]{gough_sekii}
Gough, D.O., Sekii, T. 1998,
 In: Provost J., Schmider F.X. (eds.) IAU Symp. 181: Sounding Solar and Stellar
  interior (poster volume).
 OCA \& UNSA, Nice, p.~93

\bibitem[\protect\astroncite{Gough \& Thompson}{1991}]{gough91}
Gough, D.O., Thompson, M.J. 1991,
 The inverse problem.
 In: Cox A.N., Livingston W.C., Matthews M. (eds.) Solar Interior and
  Atmosphere.
 The University of Arizona Press, Tucson, p.~519


\bibitem[\protect\astroncite{Hadamard}{1923}]{hadamard23}
Hadamard, J. 1923,
 Lectures on the Cauchy Problem in linear Partial Differential Equation, Yale
  University Press, New Haven

\bibitem[\protect\astroncite{Hansen et~al.}{1977}]{hansen77}
Hansen, C.J., Cox, J.P., Van-Horn, H.M. 1977,
 ApJ 217, 151

\bibitem[\protect\astroncite{Hansen}{1992}]{L-curve}
Hansen, P.C. 1992,
 SIAM Review 34, 561

\bibitem[\protect\astroncite{Hansen \& Mosegaard}{1996}]{PPTSVD}
Hansen, P.C., Mosegaard K. 1996,
 Numerical Linear Algebra with Applications 3(6), 513

\bibitem[\protect\astroncite{Hansen et~al.}{1992}]{MTSVD2}
Hansen, P.C., Sekii, T., Sibahashi, H. 1992,
 SIAM J. Sci. Stat. Comput. 13, 1142


\bibitem[\protect\astroncite{Kirsch}{1996}]{kirsch}
Kirsch, A. 1996,
 An Introduction to the Mathematical Theory of Inverse Problems,
  Springer-Verlag, New York

\bibitem[\protect\astroncite{Kosovichev}{1996}]{kosovichev96}
Kosovichev, A.G. 1996,
 ApJ 469, L61

\bibitem[\protect\astroncite{Kosovichev}{1998}]{kosovichev98}
Kosovichev, A.G. 1998,
 In: Provost J., Schmider F.X. (eds.) IAU Symp. 181: Sounding Solar and Stellar
  Interior (poster volume).
 OCA \& UNSA, Nice, p.~97

\bibitem[\protect\astroncite{Morel et~al.}{1997}]{morel97}
Morel, P., Provost, J., Berthomieu, G. 1997,
 A\&A 327, 349



\bibitem[\protect\astroncite{Schou et~al.}{1998}]{schou:mdi98}
Schou, J., et~al. 1998,
 ApJ,
 submitted


\bibitem[\protect\astroncite{Spiegel \& Zahn}{1992}]{spiegel92}
Spiegel, E.A., Zahn, J.P. 1992,
 A\&A 265, 106

\bibitem[\protect\astroncite{Teboul et~al.}{1998}]{teboul98}
Teboul, S., Blanc-F\'eraud, L., Aubert, G., Barlaud M. 1998,
 IEEE Trans. on Image Processing 7(3)

\bibitem[\protect\astroncite{Thompson et~al.}{1996}]{thompson:gong96}
Thompson, M.J. et~al. 1996,
 Science 272, 1300

\bibitem[\protect\astroncite{Tikhonov}{1963}]{tikhonov63}
Tikhonov, A.N. 1963,
 Sov. Maths. Doklady 4, 1035

\bibitem[\protect\astroncite{Tomczyk et~al.}{1995}]{tomc95}
Tomczyk, S., et al. 1995, Solar Phys. 159, 1
\bibitem[\protect\astroncite{Vogel \& Oman}{1996}]{vogel96}
Vogel, C.R., Oman, M.E. 1996,
 SIAM Journal of Scientific Computing 17(1)



\end{thebibliography}
\end{document}